\documentclass[aps,pra,twocolumn,nofootinbib, superscriptaddress,10pt]{revtex4-1}

\usepackage{mathdots}
\usepackage{amsfonts,amssymb,amsmath}
\usepackage[]{graphics,graphicx,epsfig}
\usepackage{amsthm}
\usepackage{csquotes}
\MakeOuterQuote{"}
\usepackage{epstopdf}

\def\identity{\leavevmode\hbox{\small1\kern-3.8pt\normalsize1}}

\newcommand{\ket}[1]{\left | #1 \right\rangle}
\newcommand{\bra}[1]{\left \langle #1 \right |}

\newcommand{\diag}{\operatorname{diag}}

\newcommand{\tr}{\operatorname{tr}}
\newcommand{\1}{\operatorname{\uppercase\expandafter{\romannumeral1}}}
\newcommand{\2}{\operatorname{\uppercase\expandafter{\romannumeral2}}}
\newcommand{\3}{\operatorname{\uppercase\expandafter{\romannumeral3}}}
\newcommand{\4}{\operatorname{\uppercase\expandafter{\romannumeral4}}}
\newcommand{\5}{\operatorname{\uppercase\expandafter{\romannumeral5}}}
\newcommand{\6}{\operatorname{\uppercase\expandafter{\romannumeral6}}}

\newcommand{\rmi}{\mathrm{i}}

\newcommand{\rmA}{\mathrm{A}}
\newcommand{\rmB}{\mathrm{B}}
\newcommand{\rme}{\operatorname{e}}

\newcommand{\caH}{\mathcal{H}}
\newcommand{\caB}{\mathcal{B}}

\renewcommand{\epsilon}{\varepsilon}

\def\eqref#1{\textup{(\ref{#1})}}
\newcommand{\eref}[1]{Eq.~\textup{(\ref{#1})}}

\def\<{\langle}  
\def\>{\rangle}  

\newcommand{\rcite}[1]{Ref.~\cite{#1}}

\begin{document}
\title{Efficient verification of bipartite pure states}

\author{Zihao Li}
\email{These authors contributed equally to this work.}
\affiliation{Department of Physics and Center for Field Theory and Particle Physics, Fudan University, Shanghai 200433, China}
\affiliation{State Key Laboratory of Surface Physics, Fudan University, Shanghai 200433, China}

\author{Yun-Guang Han}
\email{These authors contributed equally to this work.}
\affiliation{Department of Physics and Center for Field Theory and Particle Physics, Fudan University, Shanghai 200433, China}
\affiliation{State Key Laboratory of Surface Physics, Fudan University, Shanghai 200433, China}

\author{Huangjun Zhu}
\email{zhuhuangjun@fudan.edu.cn}
\affiliation{Department of Physics and Center for Field Theory and Particle Physics, Fudan University, Shanghai 200433, China}
\affiliation{State Key Laboratory of Surface Physics, Fudan University, Shanghai 200433, China}
\affiliation{Institute for Nanoelectronic Devices and Quantum Computing, Fudan University, Shanghai 200433, China}
\affiliation{Collaborative Innovation Center of Advanced Microstructures, Nanjing 210093, China}

\begin{abstract}
 We propose	practical and efficient protocols  for verifying bipartite pure states for any finite dimension, which can also be applied to fidelity estimation. Our protocols are
based on adaptive local projective measurements with either one-way or two-way communications, which are very easy to implement in practice.
They can extract the key information much more efficiently than known protocols based on tomography or direct fidelity estimation, and their efficiencies are
comparable to the best protocols based on entangling measurements.  These protocols highlight the significance of mutually unbiased bases (MUBs) and complex projective 2-designs in quantum state verification and fidelity estimation. Moreover, our protocols can be applied to the  adversarial scenario, in which states are controlled by a malicious adversary. In this case, surprisingly, one of our protocols based on local measurements is even optimal among protocols without locality restriction.
\end{abstract}

\date{\today}
\maketitle

\section{Introduction}
Bipartite and multipartite entangled states play a central role in quantum information processing, such as quantum teleportation, dense coding,  and quantum communications. They are also a focus of foundational studies on quantum entanglement and nonlocal correlations \cite{Horo09}. For real-world applications, it is crucial to  verify these states, especially pure states, with high precision, given the imperfections in the preparation devices. However, traditional approaches based on tomography are resource consuming and highly inefficient \cite{Haff05} because they extract too much unnecessary information. Even with direct fidelity estimation \cite{FlamL11}, the scaling of the number of measurements with the infidelity is suboptimal.
So far efficient verification protocols are known only for restricted states, such as stabilizer states \cite{PLM18} and hypergraph states \cite{ZhuH18}. Even for bipartite pure states, efficient protocols are known only for maximally entangled states \cite{HayaMT06, Haya08, Haya09, ZhuH19} and two-qubit  states \cite{PLM18}. This fact is both surprising and frustrating given the significance and simplicity of bipartite pure states.

In this paper we propose several simple and efficient protocols for verifying bipartite pure states, which can also be applied to fidelity estimation. To implement these protocols, it suffices to perform local projective measurements with either one-way or two-way classical communications (CCs) \cite{Owari08}. The measurement of one party may  depend on the measurement outcome of the other party. The efficiency  can be guaranteed if measurement bases of the first party are mutually unbiased \cite{Durt10, Ivano81, Woot89} or if they form a (complex projective) 2-design \cite{Zauner11,Renes04,Scott06,RoyS07,Klap05,Zhu15}. Our protocols can extract the key information---fidelity with the target state---much more efficiently than traditional approaches based on tomography and even direct fidelity estimation \cite{FlamL11}. The number of  required tests is comparable to  the best protocols based on entangling measurements, and the overhead is  at most two times even if one party can  access only two measurement settings.

Our approach can also be applied to the  adversarial scenario, in which the states are prepared by Eve. Moreover, we can construct a protocol based on local measurements that  is optimal for high-precision verification even among protocols that can access entangling measurements. As far as we know, this is the first protocol that can be applied to the adversarial scenario and achieves the optimal scaling behavior in the number of tests. This result is of key interest not only to practical applications, but also to foundational studies on the role of entanglement.

\section{Pure state verification}
In preparation for the following discussions, we first briefly review  the general framework of pure state verification \cite{PLM18}. Consider a device that is supposed to produce the target state $|\Psi\>\<\Psi|$, but actually   produces states $\sigma_1, \sigma_2, \ldots, \sigma_N$ in $N$ runs. Here we assume  either $\sigma_i=|\Psi\>\<\Psi|$ for all $i$ or $\<\Psi|\sigma_i|\Psi\>\leq1-\epsilon$ for all $i$; this assumption can be relaxed by virtue of Refs.~\cite{ZhuEVQPSshort19,ZhuEVQPSlong19}. 
To distinguish the two situations, we can perform two-outcome projective measurements on  $\sigma_i$.
Each time we choose a measurement $\{P_l,\openone-P_l\}$ with probability $p_l$ from a set of accessible measurements, where the projector $P_l$ corresponds to passing the test. To guarantee that the target state can always pass the test, the projector $P_l$ needs to satisfy the condition
$\<\Psi|P_l|\Psi\>=1$.

The above procedure is characterized  by the verification operator (or the strategy) $\Omega=\sum_{l=1}^{n} p_l P_l$.
Suppose $\bra{\Psi}\sigma \ket{\Psi} \leq 1-\epsilon$, then
the maximal probability that $\sigma$ can pass each test on average is \cite{PLM18, ZhuEVQPSlong19}
\begin{equation}
\max_{\<\Psi|\sigma|\Psi\>\leq 1-\epsilon }\tr(\Omega \sigma)=1- [1-\beta(\Omega)]\epsilon=1- \nu(\Omega)\epsilon,
\end{equation}
where  $\beta(\Omega)$ is the second largest eigenvalue of $\Omega$, and  $\nu(\Omega)=1-\beta(\Omega)$ is the spectral gap from the maximal eigenvalue. So the probability of passing $N$ tests is at most $[1- \nu(\Omega)\epsilon]^N$.  To achieve significance
level $\delta$, which means $[1- \nu(\Omega)\epsilon]^N\leq\delta$, it suffices to choose \cite{PLM18,ZhuEVQPSshort19,ZhuEVQPSlong19}
\begin{equation}\label{eq:NumberTest}
N=\biggl\lceil\frac{ \ln \delta}{\ln[1-\nu(\Omega)\epsilon]}\biggr\rceil\approx \frac{ \ln \delta^{-1}}{\nu(\Omega)\epsilon}.
\end{equation}
For high-precision verification ($\epsilon,\delta\ll1$),
the number of tests is inversely proportional to the spectral gap $\nu(\Omega)$. The optimal protocol is obtained by maximizing  $\nu(\Omega)$. If there is no restriction on the accessible measurements, then it is desirable to choose the projective measurement $\{|\Psi\>\<\Psi|,\openone-|\Psi\>\<\Psi|\}$ for each test, in which case   $\nu(\Omega)$ attains the maximum 1, and the number of tests is minimized,  $N\approx \ln \delta^{-1}/{\epsilon}$ \cite{PLM18,ZhuEVQPSshort19,ZhuEVQPSlong19}.

\section{Verification of bipartite pure states}

\subsection{Construction of tests}
Consider a bipartite system with  the  Hilbert space $\caH^{\otimes2}$ of dimension $D=d^2$, which is shared between Alice and Bob. In the Schmidt basis, any pure state in $\caH^{\otimes 2}$ can be written as $\ket{\Psi} = \sum_{j=0}^{d-1} s_j \ket{jj}$, where the Schmidt coefficients $s_j$ are arranged in decreasing order, that is, $1\geq s_0\geq s_1 \geq \cdots s_{d-1}\geq 0$, and satisfy the condition
 $\sum_{j=0}^{d-1}s_j^2 = 1$.
To verify $|\Psi\>$ efficiently, we first introduce a general method for constructing tests based on adaptive  local projective measurements. Let $\caB=\{|u_j\>\}_{j=0}^{d-1}$ be a basis on $\caH$. If Alice performs the projective measurement on this basis on the target state $|\Psi\>$ and obtains outcome $j$, then the unnormalized reduced state of Bob reads $|\tilde{v}_j\>=\<u_j|\Psi\>$. Let $|v_j\>=|\tilde{v}_j\>/\sqrt{\<\tilde{v}_j|\tilde{v}_j\>}$. Now, if Bob performs the two-outcome projective measurement $\{P_{v_j},\openone-P_{v_j}\}$ with $P_{v_j}:=|v_j\>\<v_j|$, then the target state will always pass the test (corresponding to the first outcome). So we can construct a test projector as follows
 \begin{equation}\label{eq:GenTest}
P(\caB)=\sum_{j|\<u_j|\Psi\>\neq 0}P_{u_j}\otimes P_{v_j}.
 \end{equation}

In general, the measurement of Bob depends on the measurement outcome of Alice, but this dependence can be eliminated if $|\tilde{v}_j\>$ are mutually orthogonal \cite{ZhuH19}. For example, let $\caB_0=\{|j\>\}_{j=0}^{d-1}$ be the standard basis, then
\begin{equation}\label{eq:StanTest}
P_0=P(\caB_0)=\sum_{j|s_j>0}\ket{jj} \bra{jj}.
\end{equation}
This standard test can be implemented without adaptive measurements: Both Alice and Bob perform projective measurements on the standard basis, and the test is passed if they obtain the same outcome $j$ with $s_j>0$. This test or its variants will be employed in all verification protocols  proposed in this paper.

If  $s_0=1$, then  $|\Psi\>$ is separable and can be verified efficiently using the standard test $P_0$ alone. In the rest of this paper, we assume $s_0<1$, which means  $|\Psi\>$ is entangled.  Then  each test projector based on local projective measurements has rank at least 2, so
we need at least two distinct  tests to verify $|\Psi\>$.

\subsection{Simplest verification protocols}
 Here we show that any bipartite pure state can be verified with only two distinct tests constructed from mutually unbiased bases (MUBs). Recall that two bases $\{|\psi_j\>\}_{j=0}^{d-1}$ and $\{|\varphi_k\>\}_{k=0}^{d-1}$ on $\caH$ are mutually unbiased if $|\<\psi_j|\varphi_k\>|^2=1/d$ \cite{Durt10, Ivano81, Woot89}. Let
 $\caB_1=\{\ket{u_j}\}_{j=0}^{d-1}$ be any basis that is unbiased with the standard basis $\caB_0$. A simple example is the Fourier basis $\bigl\{\sum_{k=0}^{d-1} \omega^{j k} |k\>/\sqrt{d}\bigr\}_{j=0}^{d-1}$, with  $\omega=\rme^{\rmi\frac{2\pi}{d}}$ being a primitive root of unity, but our result will not depend on the specific choice.
The test projector $P_1=P(\caB_1)$ can be computed using  \eref{eq:GenTest} above with $\ket{v_j}=M|u^*_j\>$, where $|u^*_j\>$ denotes the complex conjugate of $|u_j\>$ with respect to the standard basis, and $M:=\sqrt{d}\diag(s_0,\dots,s_{d-1})$. So we have
\begin{equation}
\tr(P_0 P_1)=\sum_{j,k=0}^{d-1}d s_k^2 |\<k|u_j\>|^2|\<k|u^*_j\>|^2=1.
\end{equation}
Let $\bar{P}_j=P_j-|\Psi\>\<\Psi|$ for $j=0,1$. 
Then we can deduce that  $\tr(\bar{P}_0\bar{P}_1)=\tr(P_0 P_1)-1=0$, which means the two projectors $\bar{P}_0$ and $\bar{P}_1$ have orthogonal supports.

If we perform the two tests $P_0$ and $P_1$ with probability $p$ and $1-p$, respectively, then the
verification operator reads $ \Omega_{\1}=pP_0+(1-p)P_1$, with
 \begin{align}
 \beta(\Omega_{\1})&=\bigl\|\bar{\Omega}_{\1}\bigl\|=\max\{p, 1-p\}\geq\frac{1}{2},
 \end{align}
 where $\bar{\Omega}_{\1}=\Omega_{\1}-|\Psi\>\<\Psi|$.
The lower bound is saturated iff $p=1/2$, in which case we have $ \Omega_{\1}:=(P_0+P_1)/2$. The corresponding spectral gap $\nu(\Omega_{\1})$ and the number  $N(\Omega_{\1})$ of tests read
 \begin{equation}
 \nu(\Omega_{\1})=\frac{1}{2},\quad N(\Omega_{\1})\approx \frac{2}{\epsilon} \ln \delta^{-1}.
 \end{equation}
Here $\nu(\Omega_{\1})$ attains the maximum among all strategies composed of two distinct test projectors based on local projective measurements according to \rcite{ZhuH19}, so $N(\Omega_{\1})$ attains the minimum among such strategies.

\subsection{Efficient protocols based on complete sets of MUBs}
When the dimension $d$ is a prime power, there exists a complete set of $d+1$ MUBs \cite{Durt10, Ivano81, Woot89}.
In this case, we can devise a more efficient protocol using  $d+1$ tests. Let $\{\caB_0, \caB_1, \ldots, \caB_d\}$ be a complete set of MUBs, where $\caB_0$ is the standard basis. Suppose we perform the test $P_0=P(\caB_0)$ with probability $p$ and each of  the other $d$ tests $P_r=P(\caB_r)$ for $r=1,2,\ldots,d$ with probability $(1-p)/d$. Then the verification operator reads
\begin{equation}\label{eq:Omega-one-way}
\Omega_{\2}=p P_0+ (1-p)\Pi,
\end{equation}
where
\begin{equation}\label{eq:Pi}
\Pi:=\frac{1}{d} \sum_{r=1}^{d} P_r=|\Psi\>\<\Psi|+ \openone\otimes\rho_{\rmB} - \sum_{k=0}^{d-1} s_k^2|kk\>\<kk|,
\end{equation}
with  $\rho_{\rmB}:=\tr_{\rmA}(|\Psi\>\<\Psi|)$. Here the second equality follows from a more general result presented in \eref{eq:OmegaC} below and proved in the Appendix. Given  that  $\bar{\Pi}=\Pi-|\Psi\>\<\Psi|$ and $\bar{P}_0$ are orthogonal, we conclude that
\begin{align}
\beta(\Omega_{\2})&=\bigl\|\bar{\Omega}_{\2}\bigl\|=\max\{p,(1-p)s_0^2\}\geq\frac{s_0^2}{1+s_0^2}.
\end{align}
The lower bound is saturated iff $p={s_0^2}/(1+s_0^2)$, in which case we have
\begin{equation}\label{eq:nuB}
\nu(\Omega_{\2})=\frac{1}{1+s_0^2},\quad N(\Omega_{\2})\approx \frac{1+s_0^2}{\epsilon} \ln \delta^{-1}.
\end{equation}
We believe that $\nu(\Omega_{\2})$ is the maximum that is achievable using local projective measurements with one-way communications. This conclusion can be proved when $|\Psi\>$ is maximally entangled \cite{HayaMT06, Haya08, Haya09, ZhuH19}.

Although we cannot construct a complete set of MUBs when $d$ is not a prime power, we can always embed $\caH$ into a larger Hilbert space $\caH'$ whose dimension is a prime power. Then we can still apply the above strategy on $\caH'^{\otimes 2}$ and achieve the same result as in \eref{eq:nuB}.

\subsection{Alternative protocols based on 2-designs}
Next, we propose an equally efficient protocol that can be applied to any bipartite pure state without enlarging the Hilbert space. Let $\{\caB_l\}^{m-1}_{l=0}$ be $m$ basis on $\caH$ and  let $\{\caB_l, w_l\}^{m-1}_{l=0}$ be a weighted set of kets with weight $w_l$ for all kets in basis $l$, where   $w_l\geq0$ and $\sum^{m-1}_{l=0}w_l=1$.
Then $\{\caB_l, w_l\}^{m-1}_{l=0}$ forms a 2-design \cite{Zauner11, Renes04, Scott06, RoyS07} if
\begin{equation}\label{eq:2-design}
\sum^{m-1}_{l=0} w_l
\sum_{|\psi_{lj}\>\in \caB_l} |\psi_{lj}\> \< \psi_{lj}|  \otimes |\psi_{lj}^*\>\<\psi_{lj}^*|=\frac{\openone +d |\Phi\>\<\Phi|}{d+1},
\end{equation}
where
$|\Phi\>=\frac{1}{\sqrt{d}}\sum_{j} |jj\>$ is the maximally entangled state. We are interested in special  2-designs in which  $\caB_0$ is the standard basis with  weight $w_0=1/(d+1)$, while  other bases are unbiased with the standard basis. Thanks to Theorem~4.1 and Proposition~4.3 in \rcite{RoyS07}, such designs can be constructed  whenever $m\geq\lceil\frac{3}{4}(d-1)^2\rceil+1$ as follows.
Each basis $\caB_l$ for $l=1,2,\ldots,m-1$ is composed of  $d$ kets of the form
\begin{equation}
|\psi_{lj}\>=\frac{1}{\sqrt{d}} \sum_{k=0}^{d-1} e^{\rmi \theta_{l j k}} |k\>,\quad
\theta_{l j k}=2\pi \Big[\frac{jk}{d}+\frac{l\tbinom{k}{2}}{m-1}\Big]
\end{equation}
for $j=0,1,\ldots, d-1$, with weight $w_l=d/[(m-1)(d+1)]$.

Given a 2-design $\{\caB_l, w_l\}^{m-1}_{l=0}$ with the desired properties,  we can devise a verification protocol by performing the test $P_0$ with probability $p$ and test $P_l=P(\caB_l)$ for $l=1,2,\ldots,m-1$ with probability $p_l=(1-p)(d+1)w_l/d$. The  verification operator turns out to be identical to $\Omega_{\2}$ in \eref{eq:Omega-one-way}, as proved in the Appendix,
\begin{equation}\label{eq:OmegaC}
\Omega_{\3}=\sum_{l=0}^{m-1} p_l P_l=p P_0+(1-p)\Pi.
\end{equation}
 Therefore, the protocol based on a 2-design is equally efficient as the one based on a complete set of MUBs (when it exists).  Again, the optimal performance is attained when $p=s_0^2/(1+s_0^2)$, in which case
we have $\nu(\Omega_{\3})=(1+s_0^2)^{-1}$ and
$N(\Omega_{\3})\approx {(1+s_0^2)} \ln \delta^{-1}/\epsilon$; cf.~\eref{eq:nuB}.

\subsection{Protocols using two-way CCs}
Up to now we have considered verification strategies $\Omega$ based on  local projective measurements with only one-way communications in which the measurement of Bob depends on the measurement outcome of Alice. Here we show that the efficiency can be improved if two-way communications are taken into account. To manifest the direction of communication, we shall use $\rightarrow $, $\leftarrow$, and $\leftrightarrow$ to denote communication from Alice to Bob, Bob to Alice, and two-way communications, respectively. Consider a strategy $\Omega^{\rightarrow}=\sum_l p_l P_l^{\rightarrow}$ built on $m$ bases $\{\caB_l\}$, where $P_l^{\rightarrow}=P(\caB_l)^{\rightarrow}$ are defined in \eref{eq:GenTest}. We can construct a new strategy by  interchanging the roles of Alice and Bob and then taking the average,
\begin{align}
 \Omega^{\leftrightarrow}&=\frac{1}{2}\Omega^{\rightarrow}+\frac{1}{2}\Omega^{\leftarrow}=\frac{1}{2}\sum_l p_l (P_l^{\rightarrow} +P_l^{\leftarrow}),
\end{align}
where $P_l^{\leftarrow}=P(\caB_l)^{\leftarrow}$ are defined according to \eref{eq:GenTest}, but with the roles of Alice and Bob interchanged. The strategy  $\Omega^{\leftrightarrow}$ is at least as  efficient as $\Omega^{\rightarrow}$ due to the  inequality $\| \bar{\Omega}^{\leftrightarrow}\|\leq (\| \bar{\Omega}^{\rightarrow}\|+\| \bar{\Omega}^{\leftarrow}\|)/2$.

Now suppose $\Omega^{\rightarrow}$ is a strategy based on a 2-design (including a complete set of MUBs) as in \eref{eq:OmegaC}. Then
\begin{equation}\label{eq:OmegaTwoWay}
\Omega_{\4}^{\leftrightarrow}=\frac{1}{2}\Omega^{\rightarrow}+\frac{1}{2}\Omega^{\leftarrow}
=pP_0+(1-p)\Pi^{\leftrightarrow},
\end{equation}
where $0\leq p<1$ and
\begin{equation}
\Pi^{\leftrightarrow}:=\frac{1}{2}\Pi^{\rightarrow}+\frac{1}{2}\Pi^{\leftarrow}
=|\Psi\>\<\Psi|+ \sum_{j\not= k}\frac{s_j^2+s_k^2}{2}|jk\>\<jk|
\end{equation}
according to  \eref{eq:Pi}. Therefore,
\begin{align}
\beta(\Omega_{\4}^{\leftrightarrow})
 &= \max\biggl\{p,(1-p)\frac{s_0^2+s_1^2}{2}\biggr\}  \geq  \frac{s_0^2+s_1^2}{2+s_0^2+s_1^2}.
\end{align}
The bound  is saturated when  $p=(s_0^2+s_1^2)/(2+s_0^2+s_1^2)$,
in which case we have
\begin{equation}\label{eq:nuTwoCC}
\nu(\Omega_{\4}^{\leftrightarrow})=\frac{2}{2+s_0^2+s_1^2},\quad N(\Omega_{\4}^{\leftrightarrow}) \approx\frac{2+s_0^2+s_1^2}{2\epsilon} \ln \delta^{-1}.
\end{equation}

\begin{figure}
	\includegraphics[width=8cm]{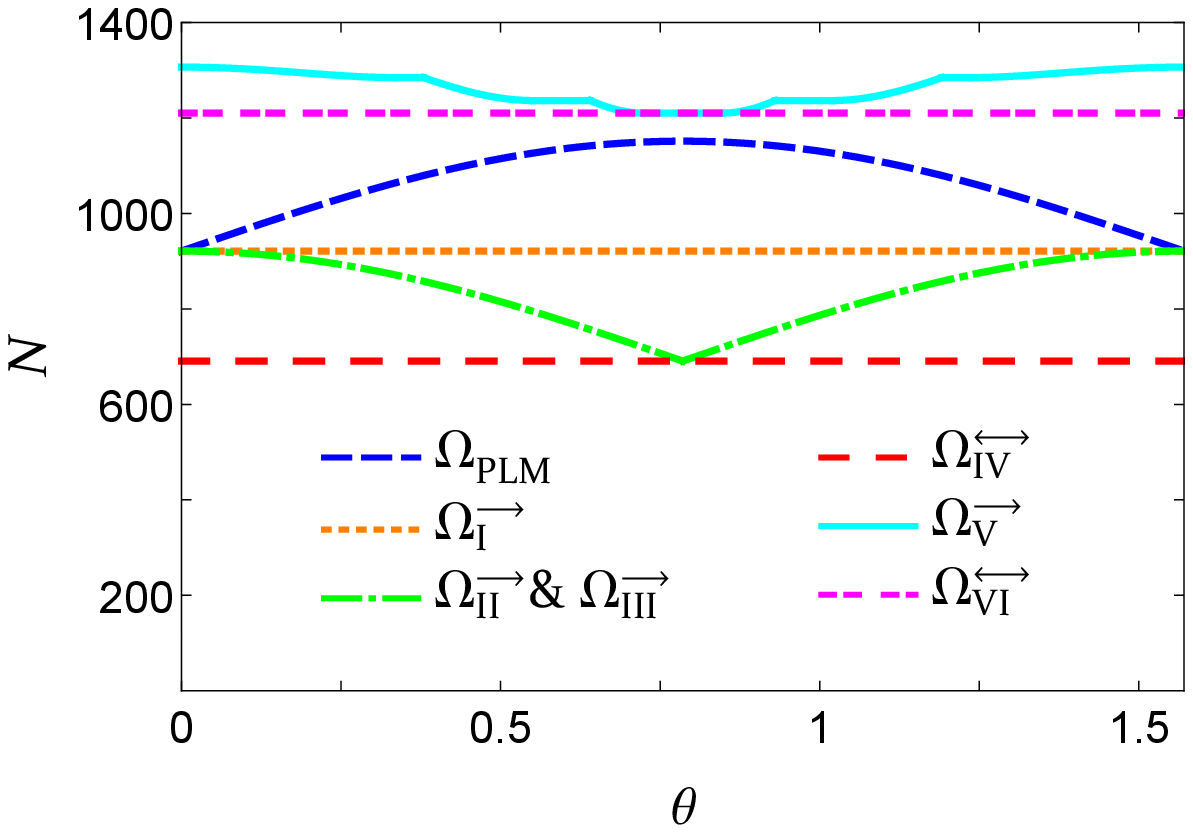}
	\caption{\label{fig:NumT}(color online). The number of tests required to verify two-qubit pure states within infidelity $\epsilon=0.01$ and significance level $\delta=0.01$. Here the target states have the form $|\Psi\>=\cos\theta\ket{00}+\sin\theta\ket{11}$. Except for the strategy $\Omega_{\mathrm{PLM}}$  introduced in \rcite{PLM18},  all other strategies are proposed in this paper. Strategies $\Omega_{\5}^{\rightarrow}$ and $\Omega_{\6}^{\leftrightarrow}$ are applied to the  adversarial scenario, while other strategies are applied to the nonadversarial scenario.
} 	
\end{figure}

In the case of two qubits ($d=2$), any entangled bipartite pure state has the form $|\Psi\>=\cos\theta\ket{00}+\sin\theta\ket{11}$ up to a local unitary transformation, where $0< \theta\leq \pi/4$. Recently, Ref.~\cite{PLM18} introduced a nonadaptive strategy $\Omega_{\mathrm{PLM}}$ with
$\nu(\Omega_{\mathrm{PLM}})=1/(2+\cos\theta\sin\theta)$. To compare with this result,  here we summarize the values of $\nu$ for the strategies proposed above
\begin{equation}
\begin{aligned}
\nu(\Omega_{\1}^{\rightarrow})&=\frac{1}{2},\quad  \nu(\Omega_{\4}^{\leftrightarrow})=\frac{2}{3}, \\
\nu(\Omega_{\2}^{\rightarrow})&=\nu(\Omega_{\3}^{\rightarrow})=\frac{1}{1+\cos^2\theta}.
\end{aligned}
\end{equation}
All four strategies are more efficient than the one introduced in Ref.~\cite{PLM18}, as illustrated in Fig.~\ref{fig:NumT}, even though the strategy $\Omega_{\1}^{\rightarrow}$ requires only two distinct tests.

\section{Adversarial scenario}
In the adversarial scenario, the state is controlled by a potentially malicious adversary \cite{HayaM15,TakeM18}. Nevertheless, we can still verify the target state using random permutations before applying a strategy $\Omega$ as in the nonadversarial scenario \cite{ZhuEVQPSshort19,ZhuEVQPSlong19}. Now the performance of  $\Omega$ will depend on smaller eigenvalues in addition to $\beta(\Omega)$. If there is no restriction on the accessible measurements, then the performance  for a given  $\beta(\Omega)$ is optimized when $\Omega$ is homogeneous, which means it has the form
\begin{equation}\label{eq:HomoStra}
\Omega=|\Psi\>\<\Psi|+\beta(\Omega)(\openone-|\Psi\>\<\Psi|).
\end{equation}
In this case, the minimum number of tests required to verify $|\Psi\>$ within infidelity $\epsilon$ and significance level $\delta$  is derived in Refs.~\cite{ZhuEVQPSshort19,ZhuEVQPSlong19}. In the high-precision limit  $\epsilon,\delta\rightarrow 0$ (assuming $\beta(\Omega)>0$), we have
\begin{equation}\label{eq:NumTestAdv}
N\approx[\beta(\Omega)\epsilon\ln \beta(\Omega)^{-1}]^{-1}  \ln \delta^{-1}.
\end{equation}
This number is minimized when $\beta(\Omega)=1/\rme$ (in contrast with the optimal value $\beta(\Omega)=0$ for the nonadversarial scenario), in which case we have
\begin{equation}\label{eq:NumTestAdvMin}
N\approx \rme\epsilon^{-1} \ln \delta^{-1}.
\end{equation}
When $\epsilon, \delta\leq0.01$, the optimal value of $\beta(\Omega)$ is usually close to $1/\rme$.

In addition to quantum state verification in the adversarial scenario, the  homogeneous strategy in \eref{eq:HomoStra} is also useful for fidelity estimation. To see this, note that the passing probability $\tr(\rho\Omega)$ of  any state $\rho$ is determined by its fidelity with $|\Psi\>\<\Psi|$, and vice versa,
\begin{equation}
\tr(\rho \Omega)=[1-\beta(\Omega)]\<\Psi|\rho|\Psi\>+\beta(\Omega).
\end{equation}
Now a small value of $\beta(\Omega)$ is favored in order to estimate the fidelity $\<\Psi|\rho|\Psi\>$ accurately \cite{ZhuEVQPSlong19}.

Here we shall devise nearly optimal homogeneous strategies by modifying $\Omega_{\3}^{\rightarrow}$ in \eref{eq:OmegaC}, which is  based on one-way communications.  Let  $s_0^2/(1+s_0^2)\leq p<1$ and  replace the test  projector $P_0$ by the following operator
\begin{equation}
Q_0=P_0'+\sum_{j\neq k}\Bigl[1-\Bigl(\frac{1}{p}-1\Bigr)s_k^2\Bigr]|jk\>\<jk|,
\end{equation}
where $P_0'=\sum_j |jj\>\<jj|$. Although $Q_0$ is not a projector, it can be realized by virtue of  local projective measurements: Both Alice and Bob  perform projective measurements on the standard basis; the test is passed with certainty if they obtain the same outcome, while with probability $1-(p^{-1}-1)s_k^2$
if they obtain outcomes $j\neq k$, respectively.
In addition, $Q_0$ can be expressed as a convex sum of local projectors. The resulting verification operator reads
\begin{equation}
\Omega_{\5}^{\rightarrow}=pQ_0+(1-p)\Pi^{\rightarrow}=|\Psi\>\<\Psi|+p(\openone-|\Psi\>\<\Psi|),
\end{equation}
which is homogeneous with $\beta(\Omega)=p$. For high-precision state verification in the adversarial scenario, the optimal choice of $p$ is  $p=\max\{\rme^{-1},s_0^2/(1+s_0^2)\}$. The resulting homogeneous strategy $\Omega_{\5}^{\rightarrow}$ is optimal if $s_0^2\leq 1/(\rme-1)$, in which case we have $\beta(\Omega)=1/\rme$. Even in the worst case $\beta(\Omega)=1/2$, the number of tests is only
$2\ln \delta^{-1}/(\epsilon\ln2)$, which is about $6\%$ more than the optimal strategy. For fidelity estimation, the optimal value of $p$ is $s_0^2/(1+s_0^2)$.

With two-way communications, we can
 devise an optimal protocol for any bipartite pure state by modifying $\Omega_{\4}^{\leftrightarrow}$ in \eref{eq:OmegaTwoWay}.
 Let  $(s_0^2+s_1^2)/(2+s_0^2+s_1^2)\leq p<1$ and
  replace the projector $P_0$ by
\begin{equation}
\tilde{Q}_0=P_0'+\sum_{j\neq k}\Bigl[1-\frac{1}{2}\Bigl(\frac{1}{p}-1\Bigr)(s_j^2+s_k^2)\Bigr]|jk\>\<jk|.
\end{equation}
Like $Q_0$, this test operator can be realized by virtue of  local projective measurements. The resulting verification operator  reads
\begin{equation}
\Omega_{\6}^{\leftrightarrow}=p \tilde{Q}_0+(1-p)\Pi^{\leftrightarrow}=|\Psi\>\<\Psi|+p(\openone-|\Psi\>\<\Psi|),
\end{equation}
which  is homogeneous with $\beta(\Omega)=p$.
If $p=1/\rme$, then $\beta(\Omega)=1/\rme$, so the number of required tests attains the minimum in \eref{eq:NumTestAdvMin} for high-precision verification in the adversarial scenario. For fidelity estimation,  the optimal value of $p$ is $(s_0^2+s_1^2)/(2+s_0^2+s_1^2)$.

\section{Summary}
By virtue of MUBs and 2-designs, we proposed a number of simple and efficient protocols for verifying general bipartite pure states. The simplest protocol requires only two distinct tests based on local projective measurements and one-way communications, yet its efficiency  is comparable to the best  protocol based on entangling measurements, and the overhead is only two times. The efficiency can be improved  further if more measurement settings or two-way communications are allowed. Moreover, our approach can be applied to the adversarial scenario, in which case one of our protocols based on local measurements is optimal even among protocols that can access entangling measurements. Our study provides a useful tool for quantum information processing with bipartite quantum states and  also offers additional insight on foundational studies on quantum entanglement.

\section*{Acknowledgment}
This work was supported by  the National Natural Science Foundation of China (Grant No. 11875110).

\emph{Note added.}
During the preparation of this paper, we
became aware of related works by Wang and Hayashi~\cite{Hayashi19},
and Yu $et\ al.$ \cite{Yu19}. Their results show that the values of $\nu(\Omega_{\2})$ in \eref{eq:nuB} and  $\nu(\Omega_{\4})$ in \eref{eq:nuTwoCC} are  optimal among protocols based on one-way communication and two-step two-way communications, respectively.

\onecolumngrid

\section*{Appendix}
\begin{proof}[Proof of \eref{eq:OmegaC}]
Note that the test projector $P_l$  can be expressed as  $P_l=(\openone \otimes M)\bigl(\sum_j|\psi_{lj}\> \< \psi_{lj}|  \otimes |\psi_{lj}^*\>\<\psi_{lj}^*|\bigr)(\openone \otimes M)$ for  $l=1,2,\ldots, m-1$, where $M=\sqrt{d}\diag(s_0,\dots,s_{d-1})$. Therefore,
\begin{align}
&\sum_{l=1}^{m-1 }w_l P_l=(\openone \otimes M)\Biggl( \sum^{m-1}_{l=1} w_l\sum_j|\psi_{lj}\> \< \psi_{lj}|  \otimes |\psi_{lj}^*\>\<\psi_{lj}^*|\Biggr)(\openone \otimes M)\nonumber\\
&=(\openone \otimes M)\Biggl(\frac{\openone +d |\Phi\>\<\Phi|-\sum_{k}|kk\>\<kk|}{d+1}\Biggr)(\openone \otimes M)=\frac{d}{d+1}\Biggl( |\Psi\>\<\Psi|+ \openone\otimes\rho_{\rmB}-\sum_{k=0}^{d-1}s_k^2|kk\>\<kk|\Biggr)=\frac{d\Pi}{d+1},
\end{align}
which implies \eref{eq:OmegaC}.
Here the second equality follows from \eref{eq:2-design}, given that $\caB_0$ is the standard basis with  weight $1/(d+1)$.
Since a complete set of MUBs is a 2-design \cite{RoyS07, Klap05, Zhu15}, \eref{eq:Pi} follows from a similar reasoning.
\end{proof}

\twocolumngrid

\end{document}